\documentclass[11pt]{article}
\usepackage[utf8]{inputenc}
\usepackage{amsmath}
\usepackage{amssymb}
\usepackage{mathrsfs}
\usepackage{array}
\usepackage{graphicx}
\usepackage[margin=1in]{geometry}
\usepackage{setspace} 
\usepackage{natbib}
\usepackage{comment}
\newtheorem{theorem}{Theorem}
\usepackage[noblocks]{authblk}
\usepackage{hyperref}
\usepackage{multicol}
\usepackage{caption}
\usepackage{subcaption}

\newcommand{\beq}{ \begin{equation}}
\newcommand{\eeq}{ \end{equation}}
\newcommand{\beqn}{ \begin{eqnarray}}
\newcommand{\eeqn}{ \end{eqnarray}}

\usepackage{caption}
\begin{document}

\title{\textbf{A new mixture model for spatiotemporal exceedances with flexible tail dependence}}
\author[1]{Ryan Li}
\author[1,2]{Emily C. Hector}
\author[1]{Brian J. Reich}
\author[3]{Reetam Majumder}
\affil[1]{North Carolina State University}
\affil[2]{University of Michigan}
\affil[3]{University of Arkansas}
\date{}
\maketitle

\abstract{We propose a new model and estimation framework for spatiotemporal streamflow exceedances above a threshold that flexibly captures asymptotic dependence and independence in the tail of the distribution. We model streamflow using a mixture of processes with spatial, temporal and spatiotemporal asymptotic dependence regimes. A censoring mechanism allows us to use only observations above a threshold to estimate marginal and joint probabilities of extreme events. As the likelihood is intractable, we use simulation-based inference powered by random forests to estimate model parameters from summary statistics of the data. Simulations and modeling of streamflow data from the U.S. Geological Survey illustrate the feasibility and practicality of our approach. 
}

\noindent%

\maketitle

\section{Introduction}\label{sec1}

Flooding from extreme streamflow (water flow through rivers and streams) contributes to the more than \$179 billion annual cost in the United States \citep{JEC}. Flooding from streamflow can be mitigated through civil engineering solutions at appropriate locations along rivers and streams. For this reason, accurately and precisely modeling extreme streamflow events at high temporal resolution, such as daily streamflow maxima, has the potential to substantially limit the economic and human impacts of flooding. Extreme streamflow, however, is difficult to model because the mechanics of flooding are complex, and extreme events are by definition rare. 

Spatial extreme value analysis is a useful tool for modeling extreme streamflow because it allows the borrowing of information across sites to improve the accuracy and efficiency of estimates, and can allow extrapolation to unobserved sites. Max stable processes (MSPs) are a limiting class of models for block maxima (maximum values evaluated over a period of time) \citep{smith1990max, tawn1990modelling, schlather2002models,kabluchko2009stationary, wadsworth2012dependence, reich2012hierarchical}, making them a natural model for extreme streamflow measured at $n$ locations. 
MSPs are typically used to model block maxima of spatio-temporal processes (e.g. annual maximum streamflow) across several time points (e.g. years) that are treated as independent replicates. This assumption of independence is violated at higher temporal resolutions, such as daily streamflow maxima, where a single flood event can be observed over multiple days. 
Further, a well-known characteristic of MSPs are their \emph{asymptotic dependence} \citep{huser2025modeling}, wherein the conditional probability of observing an extreme event given another extreme event is strictly greater than zero or, mathematically, $\chi = \lim_{y \to \infty} P(Y_2 > y | Y_1 > y) > 0$. In practice, many environmental processes show decreasing spatial dependence in the tail of the distribution as events become more extreme. Interestingly, this does not guarantee asymptotic independence. Selecting a model with an incorrect tail dependence structure will result in poor extrapolation into the joint upper tails \citep{ledford1996statistics, davison2013geostatistics}.    

Models that can capture both asymptotic dependence \emph{and} asymptotic independence are appealing alternatives to the restrictive tail dependence structure of the MSP. Examples include the max-mixture model of \citet{wadsworth2012dependence} and the copula-based mixture models in, e.g., \citet{huser2019modeling} and \citet{majumder2024modeling}. These papers specify a copula for the dependence structure separately from the marginal models. \citet{wadsworth2012dependence} specify the copula as the maximum of two marginally Fr\'{e}chet random variables, one with asymptotic dependence and one with asymptotic independence, while \citet{huser2019modeling} and \citet{majumder2024modeling} model the copula as a linear combination of an asymptotically dependent and an asymptotically independent process, to allow for flexible extremal dependence between two spatial locations. All three methods assume temporal independence, which makes them unsuitable for analyzing data observed at high temporal resolutions. 

Recent work in this area aims to extend spatial extremes analysis to include temporal dependence. For example, \citet{bortot2024model} extend the model in \citet{huser2019modeling} to incorporate temporal dependence by modeling observations at each site over time with an autoregressive Gaussian time series; however, their model is limited to temporal asymptotic independence. \citet{DELLORO2025100916} extend the \citet{huser2019modeling} model by allowing the mixture components to vary between temporal asymptotic dependence and independence, which allows for flexibility in temporal asymptotic properties for the model. This approach, however, requires an additional step to select the appropriate asymptotic dependence structures of the mixture components rather than letting the structure be chosen by the data. 
\citet{MAUMEDESCHAMPS2024100860} propose a max-mixture model with temporally varying mixture proportions which allows the model to vary between spatial asymptotic dependence and independence over time, but it cannot model temporal asymptotic dependence.   

Exact inference for spatial extremes models in general and MSPs in particular pose significant computational challenges. Evaluating the density associated with a censored MSP is computationally intractable when $n$ is even moderately large \citep{schlather2002models,wadsworth2012dependence,wadsworth-2015a}. For general MSPs, \citet{Castruccio-etal} found full inference limited to $n=13$ locations, while \citet{Huser-etal} proposed an expectation-maximization algorithm for an MSP at $n=20$ locations, with a computation time of $19.8$ hours. Furthermore, exact inference is impossible for several variants of the scale-mixture model of \citet{huser2019modeling}; this includes the models of \citet{majumder2024modeling} and \citet{DELLORO2025100916} which are closely related to this work. There are two common approaches to address the intractability of spatial extremes models. The first involves the use of a composite likelihood approximations \citep{Padoan-Ribatet-Sisson,Huser-Stein-Zhong-2022}. Composite likelihood suffers from  statistical inefficiency for large $n$ \citep{Huser-Davison-Genton}, finite-sample bias when using all pairs of observations \citep{Sang-Genton, wadsworth-2015a, Castruccio-etal}, and computational challenges posed by computing likelihoods at all $O(n^2)$ pairs. Simulation-based inference (SBI) has emerged as a popular alternative. SBI leverages the fact that while spatial extremes likelihoods might be challenging to evaluate, simulation from these processes is relatively straightforward. Broadly speaking, SBI can be categorized into methods to learn the likelihood \citep[e.g.,][]{majumder2024modeling,MAJUMDER2023100755,WALCHESSEN2024100848}, and methods to learn the posterior of parameters \citep[e.g.,][]{gerber2021fast,Banesh2021,Sainsbury-Dale02012024}; we refer readers to \citet{zammit-mangion-2025} for a recent review. For parameter estimation in particular, SBI proceeds by first generating a dense surface of plausible parameter values, and then generating data from the process for each set of parameter values. A predictive model (usually a deep learning model) is then trained to learn a mapping from a set of carefully chosen summary statistics of the dataset to the parameters. This functions as a surrogate estimator that, when supplied with a dataset, can output the parameter values that the dataset is generated from. In spatial statistics, the approach was originally used to learn the parameters of the covariance matrix of a Gaussian process \citep{gerber2021fast,Banesh2021}; it has since been used for parameter estimation in a wide range of models of spatial extremes \citep{lenzi2023neural,Sainsbury-Dale02012024,Richards2024,hector2024whole,DELLORO2025100916,Sainsbury-Dale03072025}.

Our modeling framework has a product-sum correlation structure similar to the one outlined in \citet{de2001estimating}, which is a class of stationary spatio-temporal covariance functions used in the context of Gaussian models that are expressed as a weighted sum of spatial, temporal, and a product of spatial and temporal covariance functions. Gaussian models are not suitable for analysis of extreme data, as they are asymptotically independent and may not capture extremal dependence of observations. This structure is useful, however, due to its flexibility and guaranteed strict positive-definiteness \citep{de2011strict}. It has been used in analysis of stream network data \citep{bachrudin2023spatio} which makes it a natural approach to expand on for this application.

\paragraph{Our contributions:} We propose a new model and inferential framework that incorporates the flexibility of models proposed in \citet{huser2019modeling} and \citet{majumder2024modeling} with respect to extremal dependence, but that also incorporates a flexible model for temporal dependence. Relative to the two aforementioned works, our key idea is to introduce additional mixture components to the model that each represent asymptotic dependence or independence across space and/or time. Our model can therefore flexibly capture any of the four asymptotic (in)dependence regimes (space/time and independent/dependent), thereby improving prediction of daily streamflow maxima at unobserved sites and time points. Put into practice, our approach can improve the forecasting of extreme streamflow events at unobserved sites and present targeted locations and times for developing engineering solutions to prevent flood damage.

We use a peaks-over-threshold model for the marginal distribution of the data. This bears some resemblance to the approaches in \citet{Reich-Shaby-Cooley, Huser-Davison-2014, hector2024distributed}, where MSPs are used to model peaks over a threshold using a censored likelihood. Censoring observations below a threshold prevents non-extreme values from biasing our estimated model for the extremes. A major drawback of using more data and our more complex model is computational tractability. To overcome this computational difficulty, we use random forests \citep{Breiman2001} trained on summary statistics of our data to estimate parameters of our model. Notably, we use empirical $\chi_u$ values for pairs of observations as our predictors.
Our use of random forests is a departure from the norm for SBI which has largely employed deep learning up until now. Deep learning is the popular choice due to its universality, but random forests are computationally cheaper and have fewer tuning parameters, making them preferable if there is minimal loss of accuracy and precision. 

The remainder of the paper is organized as follows. Section \ref{s:background} provides background on MSPs and tail dependence measures, and summarizes the mixture models of \citet{huser2019modeling} and \citet{majumder2024modeling}. Section \ref{s:model} describes our spatio-temporal model. Section \ref{s:estimation} describes our proposed inferential framework for simulation based inference using random forests. Section \ref{s:simulations} investigates model performance in terms of goodness-of-fit and inference through a simulation study. Section \ref{s:streamflow} analyzes daily streamflow maxima from U.S. Geological Survey (USGS) Hydro-Climatic Data Network (HCDN) stations to identify trends in extreme streamflow over the past 60 years using the proposed methods. Additional theoretical results are available in the Appendix. Data and code for the paper can be found at \texttt{https://github.com/rli2000/SpatialExtremeMixture}.

\section{Background}\label{s:background}

\subsection{$\chi$-coefficient}\label{s:chi}

Spatial and temporal dependence is often measured by the $\chi$ coefficient \cite{Coles2001}.  Let $Y(s,t)$ denote a random process $Y$ observed at spatial locations $s$ and time points $t$. If we denote $q(\tau; s,t)$ as the $\tau$ quantile of $Y(s,t)$, then the $\chi$ coefficient between $Y(s,t)$ and $Y(s',t')$ is 
\begin{eqnarray}\label{e:chi}
    \chi_{\tau}(s,t,s',t') &=& \mbox{Prob}\left[Y(s,t)>q(\tau; s,t)|Y(s',t')>q(\tau; s',t')\right],\\
    \chi(s,t,s',t') &=& \lim_{\tau\rightarrow 1}\chi_{\tau}(s,t,s',t').
\end{eqnarray}
For an isotropic process, we can write $\chi(h_S,h_T)$ for $h_S=||s-s'||$ and $h_T=||t-t'||$.  The isotropic process $Y$ is asymptotically dependent in space if $\chi(h_S,0)>0$; asymptotically dependent in time if $\chi(0,h_T)>0$ for all $s$; and asymptotically dependent in space and time if $\chi(h_S,h_T)>0$ for all $h_S$ and $h_T$. 

\subsection{Max-stable and inverted max-stable process}

A stochastic process $\{ Y_t, t \in T\}$, where $T$ is an index set, is a max-stable process if, for some constants $A_{Nt} > 0, B_{Nt}: (N \ge 1, t \in T)$, and $Y_t^{(n)}: n = 1, \cdots, N$ independent, identically distributed copies of $Y_t$, such that $Y^*_t =  \{\max_{1\leq n\leq N}Y_t^{(n)} - B_{nt}\}/A_{nt} $, then $\{Y^*_t: t \in T\}  \overset{\mathrm{d}}{=} \{ Y_t, t \in T\}$  \citep{smith1990max}. This definition requires the marginal distribution of max-stable processes to be a generalized extreme value distribution.
Max-stable processes are the infinite-dimensional generalization of multivariate extreme value theory \citep{smith1990max}, so they are a natural model for spatial extremes, which are infinite-dimensional extreme values.

A finite parametrization for the general max-stable process does not exist, so models are derived from a spectral representation \citep{de1984spectral}. Brown-Resnick processes are one such parametrized subclass of max-stable processes. A Brown-Resnick process can be defined as $Z(s) = \sup_{i \in \mathbb{N}} W_i(s) / T_i$, where $T_1 < T_2 < \cdots $ are points of a Poisson process with rate 1, and $W_i(s)$ are copies of a random process $W(s) = \exp{\{\epsilon(s) - \gamma(s)\}}$, where $\epsilon(s)$ is a Gaussian process with semivariogram $\gamma(s)$ and $\epsilon(0) = 0$ almost surely \citep{brown1977extreme, huser2013composite}. The multivariate CDF for sites $s \in D$ is $P\{Z(s) < z(s), s \in D\} = \exp[ -V_D\{ z(s) \} ]$, where $V_D\{ z(s) \} = E\{ \sup_{s \in D}W(s)/z(s) \}$, has explicit formula for $|D|\in \{1,2\}$. This likelihood becomes computationally intractable for more than a few sites, requiring approximations such as composite likelihood or Vecchia approximation \citep{huser2013composite, Huser-Stein-Zhong-2022}.
Other families of max-stable models include the Schlather model \citep{schlather2002models}, and the extremal-t model \citep{brown1977extreme}.
Finally, if $Z(s)$ is a max-stable process with marginal CDF $G_s(x)$, then $W(s) = 1/G_s\{ Z(s) \}$ has an inverted max-stable copula with Pareto margins \citep{wadsworth2012dependence, ledford1996statistics}. This process has positive association between pairs of observations, but asymptotic dependence. 

\subsection{Regression Trees and Random Forests}
Regression trees are a machine learning tool that sequentially splits the training data into groups in a binary tree structure \citep{Breiman2017-bj}. At each step, the training data is split to minimize $\sum_t \sum_{y_n \in t} \{ y_n - \overline{y}(t) \}^2$, the sum of squared deviations across the two new groups. This process is repeated until the size of the resulting groups reaches a predefined lower bound, and then a pruning procedure is implemented to reduce the complexity of the tree. To predict the response value of a new observation, the leaf node corresponding to the new observation's predictor values is found, and the average response of training data in that node is taken as the prediction.

A random forest is an ensemble predictor consisting of $k$ regression trees, $\{h(x,\Theta_i) : i = 1 \cdots,k\}$, where $\{\Theta_i\}$ are independent, identically distributed random vectors, that averages the $k$ predictions from the trees \citep{Breiman2001}. Each tree is grown from a subset of the response vector, $\Theta_i$ that is sampled from $Y$ with replacement. This approach is popular for high-dimensional regression due to its power and interpretability \citep{quantregforest}.

The tuning parameters for random forests are the number of trees $k$, and the length of the random vectors $\Theta_i$, as well as the tuning parameters for the individual random tree predictors, the minimum group size.

\section{Spatiotemporal mixture model}\label{s:model}
\subsection{Tail model}\label{s:stmixture}

We propose a model that extends the spatial mixture model in \citet{huser2019modeling} and \citet{majumder2024modeling} to the spatiotemporal setting to allow for asymptotic dependence or independence in space and/or time depending on model parameters.  Let $i \in \{1, \cdots, n\}$ index independent replicates (e.g., years), $t\in \mathcal{T}$ index correlated time points (e.g., days within year), and $s\in \mathcal{S} \subseteq {\cal R}^2$ denote the spatial location.  The observations $Y_i(s,t)$ for $i \in \{1, \cdots, n\}$, $t=\{1, \ldots, T\}$, $s\in \{s_1, \ldots, s_m\}$, are instances of the random process $Y(s,t)$ at time $t$ and location $s$, whose joint distribution we describe below. We denote the marginal distribution of $Y(s,t)$ by $F(y; s,t)$. 
In our analysis of the streamflow data in Section \ref{s:streamflow}, we assume that the upper tail of $F(\cdot; s,t)$ can be approximated by a generalized Pareto distribution (GPD) with parameters that vary across space and time; censoring is discussed in Section \ref{s:estimation}.  

To model spatiotemporal dependence, we define a latent process $X(s,t)$ that is a transformation of the response, $Y(s,t) = G\{ X(s,t) ; s,t\}$, where the transformation function $G(\cdot; s,t)$ is defined below.  Note that since $X$ and $Y$ are one-to-one, they share asymptotic dependence properties as measured by $\chi(h_S,h_T)$. 
We model the latent process as a convex combination of four processes,
\begin{equation}\label{e:Xmixture}
    X(s,t) = \lambda_1 R_{ST}(s,t) + \lambda_2 R_{S}(s,t) + \lambda_3 R_{T}(s,t) + \lambda_4 W(s,t),
\end{equation}
where $R_{ST}$, $R_S$, $R_T$ and $W$ each marginally follow standard exponential distributions and the mixture weights $\lambda = (\lambda_1,...,\lambda_4)$ satisfy $\lambda_k\ge0$ and $\sum_{k=1}^4 \lambda_k = 1$. The four processes correspond to four distinct asymptotic regimes: $R_{ST}$ is asymptotically dependent over space and time, $R_S$ is asymptotically dependent over space but not time, $R_T$ is asymptotically independent over time but not space and $W$ is asymptotically independent over space and time.   

Specifically, $R_{ST}$, $R_S$ and $R_T$ are taken as max-stable processes and $W$ is an inverted max-stable process, all marginally transformed to have standard exponential distributions.  The spatiotemporal process $R_{ST}$ is a Brown-Resnick process \citep{brown1977extreme,kabluchko2009stationary} with spatiotemporal  correlation function $$\exp \{-(h_S/\rho_S)^\alpha - (h_T/\rho_T)^\alpha \},$$ where $h_S$ is the distance between two locations and $h_T$ is the distance between two time points.  The  spatial Brown-Resnick process $R_S$ has spatial variogram $ (h_S/\rho_S)^\alpha $ and is completely independent over time.  Similarly, the temporal Brown-Resnick process $R_T$ has temporal variogram $(h_T/\rho_T)^\alpha $ and is independent over space.  Finally, $W$ is an inverted Brown-Resnick process \citep{wadsworth2012dependence} with spatial variogram $(h_S/\rho_S)^\alpha + (h_T/\rho_T)^\alpha $, which is asymptotically independent over space and time.

The marginal distribution of $X(s,t)$ for all $s$ and $t$ is the hypoexponential distribution with parameters $(\lambda_1,\lambda_2,\lambda_3,\lambda_4)$ and cumulative distribution function

\begin{equation}\label{e:hypoexpoecf} 
H_{\lambda}(x) = 
1 - \Big( \prod_{k = 1} ^4 \lambda_k^{-1} \Big) \sum_{j = 1}^4 \Big[\Big\{-\lambda_j^{-1} \prod_{k \ne j} (\lambda_k^{-1} - \lambda_j^{-1} )\Big\}^{-1}e^{-x/\lambda_j} \Big],
\end{equation}
for distinct $\lambda_1, \cdots, \lambda_4$.
We therefore use $G(x; s,t) = F^{-1}\{H_{\lambda}(x); s,t\}$ to relate the latent process $X(s,t)$ to the response $Y(s,t) = G\{X(s,t); s,t\}$.  By construction, the process $Y(s,t)$ has marginal distribution function $F(\cdot; s,t)$ and inherits the spatiotemporal dependence properties of $X(s,t)$.

To simplify the model, we assume the spatial and temporal range parameters $\rho_S$ and $\rho_T$ are constant across all four mixture components, and the smoothness parameter is fixed at $\alpha=1$. Since the interpretation of $\rho_S$ is the same across all four mixture components, we can treat it as the overall spatial range for the process $X(s,t)$, and likewise for $\rho_T$.   The gives spatial dependence parameters $\theta = (\lambda_1,\lambda_2,\lambda_3,\lambda_4,\rho_S,\rho_T)$. Using this model, we can capture any combination of spatial and temporal asymptotic dependence regimes by varying $\theta$, as shown in Section \ref{s:AD}. 

\subsection{Asymptotic tail dependence}\label{s:AD}

We study the asymptotic dependence structure of a simplified version of the process in equation \eqref{e:Xmixture} given by
\begin{equation}\label{e:Xmixturesimple}
    X(s,t) = \lambda_1 R_1 + \lambda_2 R_{2}(t) + \lambda_3 R_{3}(s) + \lambda_4 R_4(s,t),
\end{equation}
where $R_{1}$, $R_2(t)$, $R_3(s)$ and $R_4(s,t)$ all follow mutually independent standard exponential distributions. This simplifies derivations but maintains the key dependence properties of each component; for example, $R_2(t)$ is constant in $s$ and thus dependent across space but not time. We further assume that the $\lambda_1,...,\lambda_4$ are distinct and denote $j = \arg \max_k \lambda_k$ as the dominant term. The following Theorem \ref{t:ad} shows that the simplified model of \eqref{e:Xmixturesimple} is sufficiently flexible to account for (a) spatial, (b) temporal and (c) spatiotemporal dependence, and that inference on the dominant term $j$ provides a way to distinguish between these modes of dependence.  

\begin{theorem}\label{t:ad}
Under the model in \eqref{e:Xmixturesimple}:
\begin{enumerate}
\item[(a)] $X(s,t)$ and $X(s',t)$ are asymptotically dependent if and only if $j\in\{1,2\}$,
\item[(b)] $X(s,t)$ and $X(s,t')$ are asymptotically dependent if and only if $j\in\{1,3\}$,
\item[(c)] $X(s,t)$ and $X(s',t')$ are asymptotically dependent if and only if $j=1$.
\end{enumerate}
\end{theorem}
A proof is provided in Appendix A. 

\begin{figure}
\caption{Empirical $\chi$ plots for four $\lambda$ settings for $X(s,t), X(s^\prime,t)$ (left), $X(s,t), X(s^\prime,t^\prime)$ (middle) and $X(s,t), X(s,t^\prime)$ (right).}\label{f:model:chi}\centering 
\includegraphics[width=0.99\textwidth,page=1, trim = {0 6cm 0 6cm}]{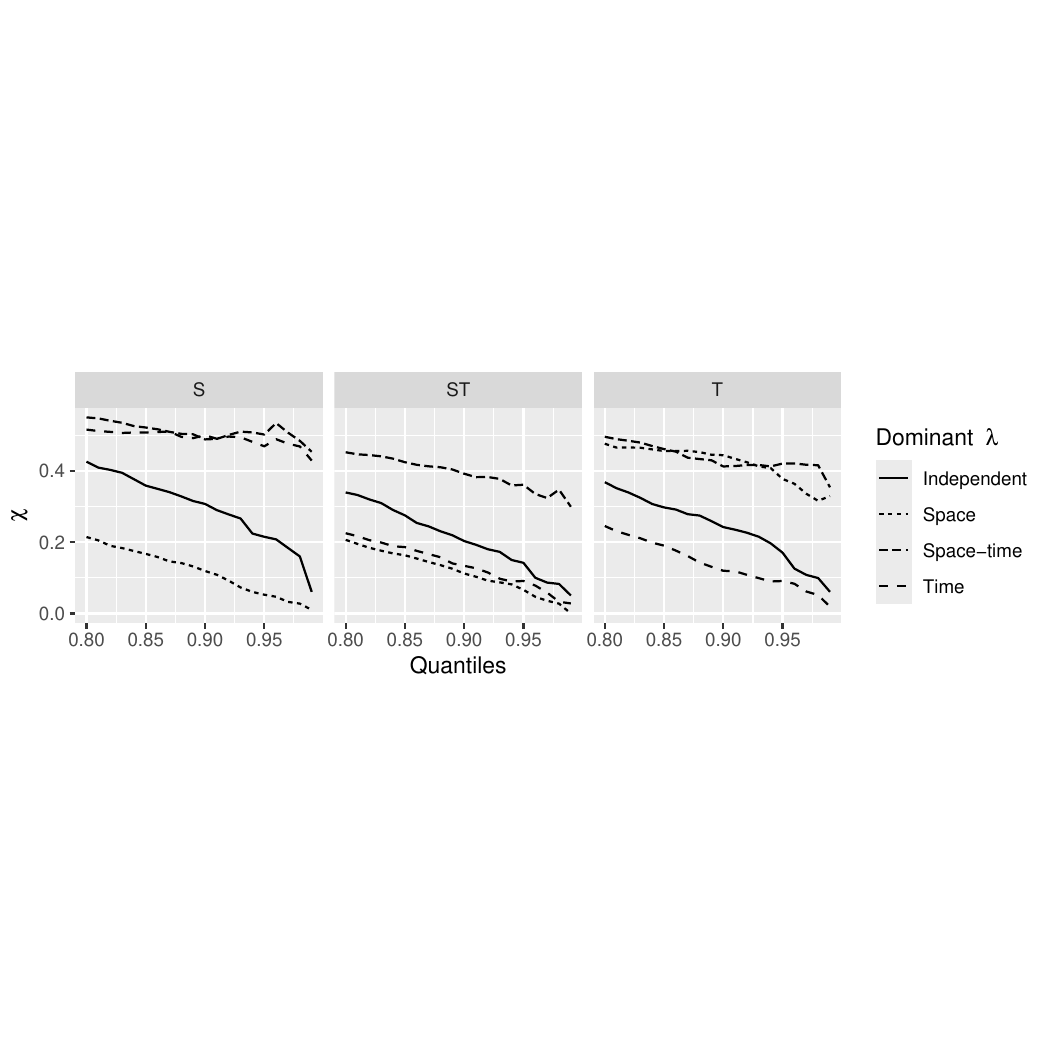}
\end{figure}

While the derivations for the full model in \eqref{e:Xmixture} are cumbersome, we explore results analogous to Theorem \ref{t:ad} for the full model graphically in Figure \ref{f:model:chi}. We generate 10,000 draws from the mixture model with parameters $\rho_S = \rho_T = 0.2$ at four spatio-temporal locations: $(0,0,0), (0.8,0,0), (0,0,0.4), (0.8,0,0.4)$. Each model has dominant $\lambda_j = 0.4$, and remaining terms $\lambda_k \approx 0.2$, with minor perturbations to maintain distinct values. The empirical results support the conclusion of the theorem for the model in \eqref{e:Xmixture}; for two spatial sites, $\chi_u \to 0$ when $j \in \{3,4\}$ and otherwise $\chi_u$ approaches a larger value, indicating likely asymptotic dependence. Similar results follow for two points separated in time, and separated in both space and time.

\section{Parameter estimation}\label{s:estimation}

\subsection{Marginal parameter estimates}

We first transform the data to standard uniform margins before estimating the dependence parameters $\theta$. We do this by modeling the marginal distribution of the data and performing a probability integral transformation. Since we are modeling with a points-above-threshold approach, we model the marginal distribution of our data with a Generalized Pareto distribution with threshold taken to be the $q_0$ marginal quantile for each spatial location. We use maximum likelihood estimation (MLE) to estimate marginal parameters, with one shared shape parameter and separate range parameters for each spatial location.  The observations $Y_i(s,t)$ are then transformed to $U_i(s,t)\in[0,1]$ using the fitted distribution function; observations below the threshold can be set to any value below the threshold because, as described below, they do not affect the estimation of $\theta$. 

 \subsection{Dependence parameter estimates}

One challenge in using full likelihood inference to estimate our model is to evaluate the likelihood. Indeed, the max-stable processes in our mixture model have a computationally intractable full likelihood for more than a small number of observations. This computational problem is exacerbated by censoring of observations in our peaks over threshold analysis. As it is straightforward to simulate data from the model, we instead use SBI for parameter estimation.

First, we generate $S$ draws from the prior distribution of the parameters. To work in an unbounded domain, we apply the following transformation: $\eta_1 = \log(\lambda_2/\lambda_1), \eta_2 = \log(\lambda_3/\lambda_1), \eta_3 = \log(\lambda_4/\lambda_1), \eta_4 = \log{\rho_S}$ and $\eta_5 = \log{\rho_T}$. The prior distributions are $\eta_j\sim \mbox{Normal}(m_j,s_j^2)$, $j=1, \ldots, 5$.  For each of the $S$ parameter sets, we draw a dataset from the model using the same spatiotemporal locations and number of independent replicates as the dataset to be analyzed. The datasets are then reduced to summary statistics, $Z$, as described below.  We then use the $S$ samples to train a random forest model for each parameter, ${\hat \eta}_j = {\hat f}_j(Z)$, $j=1, \ldots, 5$.  To estimate a parameter given a real dataset, we simply compute the summary statistics for the real dataset and pass them through the fitted model ${\hat f}_j$, $j=1, \ldots, 5$.  For uncertainty quantification, we use non-parametric bootstrap (i.e., replications sampled with replacement) to derive both estimates of standard errors and parameter confidence intervals. 

The summary statistics are derived from empirical versions of the $\chi$ coefficients in (\ref{e:chi}).  We compute ${\hat \chi}_{\tau}(s,t,s',t')$ for thresholds $\tau\in\{q_1,...,q_m\}$ across all pairs of spatiotemporal locations. Of course, the thresholds $q_j$ should be at least as large as the marginal threshold, $q_0$. Separately by $\tau$, the empirical $\chi$ coefficients are binned over spatial and temporal distance. We bin observations by every 0.1 increment in spatial or temporal distance (with each spatiotemporal dimension scaled to $[0,1]$), with five spatial bins and three temporal bins, and then smoothed using a 2D B-spline to reduce noise in the model inputs. In the simulation study of Section \ref{s:simulations} we use $m=2$, $q_1=0.5$ and $q_2=0.9$, but in practice this will require tuning.  
Additionally, we consider the statistic $\text{corr}\{ U(s,t), U(s',t') | U(s,t) > q_0, U(s',t') > q_0\}$. Since we consider only values above $0.5$ in the correlation calculation, we effectively censor the data at the $0.5$ quantile level because data below the threshold will not affect our estimates. Given the summary statistics, the regression functions ${\hat f}_j$ are fitted using the $S$ parameters/summary statistic pairs using random forest with default settings in the \texttt{quantregForest} package \citep{quantregforest}, separately across parameters $\eta_1, \cdots, \eta_5$.  

Finally, we transform from $\eta$ to $\theta$ using the inverse transformations: $\lambda_1 = 1/\{\sum_{i=0}^3 \exp(\eta_i)\}$, $\lambda_2 = \exp(\eta_1)/\{\sum_{i=0}^3 \exp(\eta_i)\}$, $\lambda_3 = \exp(\eta_2)/\{\sum_{i=0}^3 \exp(\eta_i)\}$, $\lambda_4 = \exp(\eta_3)/ \{\sum_{i=0}^3 \exp(\eta_i)\}$, $\rho_S = \exp{\eta_4}$, and $\rho_T = \exp{\eta_5}$, where $\eta_0 = 0$. We apply this transformation to the estimates before computing bootstrap confidence intervals.

\section{Simulation experiments} \label{s:simulations}

We investigate the properties of our model and estimation procedure through numerical experiments on simulated data. The spatio-temporal domain $\mathcal{T} \times \mathcal{S}$ is a three-dimensional unit cube and observations are sampled on a uniform grid of dimension $5\times 5 \times 5$. That is, $t\in \{1, \ldots, 5\}$ and $s \in \{ (a,b) \}_{a, b=1}^5$ such that $T = 5, m = 25$. We construct four settings where $\lambda_i$ is the dominant term for each $i \in \{1,2,3,4\}$. In the setting where $\lambda_i$ is dominant, $\lambda_i = 0.5, \lambda_j \in \{0.15,0.17,0.18\}$ for $j \ne i$; the remaining $\lambda_j$ are assigned values based on index order. In all scenarios and settings, the spatial and temporal range values are $\rho_S = \rho_T = 0.4$, and the marginal shape and scale parameters are $0.2$ and $1$, respectively. The scale parameters are estimated individually for each of the 25 sites, but for brevity only $\sigma_1$, the scale at location 1, is reported. For each setting, we run 100 simulations. In all scenarios, the number of independent replicates is $n = 100$. Parameter estimation is carried out using the random forest approach described in Section \ref{s:estimation}. 

Table \ref{t:sim} shows that our model estimates the $\lambda$ parameters very well. The bias is generally low and the coverage is near the nominal level and in these synthetic cases we are able to consistently identify the asymptotic regime. Additionally, the range parameters $\rho_S$ and $\rho_T$ are reasonably estimated in most cases, although bias and low coverage is observed when data are generated with strong asymptotic dependence. 

\begin{table}[htbp]
\centering
 \begin{subtable}[h]{\textwidth}
        \centering
        \begin{tabular}{rrrrrrr}
Parameter & True & Mean & Bias & RMSE & SE & Coverage \\ 
  \hline
  $\lambda_1$ & 0.500 & 0.467  &  0.091  &  0.008 & 0.085 & 0.84   \\   
  $\lambda_2$ & 0.150 & 0.122  &  0.059  &  0.003 & 0.052 & 0.89\\ 
  $\lambda_3$ & 0.170 & 0.178  &  0.045  &  0.002 & 0.045 & 1.00\\
  $\lambda_4$ & 0.180 & 0.233 &   0.116  &  0.013 & 0.103 & 0.91\\ 
  $\rho_S$ & 0.400 & 0.581  &  0.213  &  0.045 & 0.112 & 0.75\\ 
  $\rho_T$ & 0.400 & 0.534  &  0.156  &  0.024 & 0.081 & 0.77\\ 
  $\xi$ & 0.200 & 0.163 & -0.036 & 0.172 & 0.119\\ 
  $\sigma$ & 1.000 & 1.125 & 0.125 &  0.101 &  0.095
\end{tabular}
       \caption{ Asymptotic spatiotemporal dependence with $\lambda_1=0.5$}
    \end{subtable}
    \vfill
    \begin{subtable}[h]{\textwidth}
        \centering
       \begin{tabular}{rrrrrrr}
Parameter & True & Mean & Bias & RMSE & SE & Coverage \\ 
  \hline
  $\lambda_1$ & 0.150 & 0.127  &  0.034  &  0.001 & 0.024 & 0.88\\ 
  $\lambda_2$ & 0.500 & 0.446  &  0.074  &  0.005 & 0.051 & 0.97\\ 
  $\lambda_3$ & 0.170 & 0.223  &  0.080  &  0.006 & 0.060 & 0.97\\ 
  $\lambda_4$ & 0.180 & 0.204  &  0.083  &  0.007 & 0.080 & 0.94\\ 
  $\rho_S$ & 0.400 & 0.419  &  0.049  &  0.002 & 0.046 & 0.98\\ 
  $\rho_T$ & 0.400 & 0.250  &  0.163  &  0.027 & 0.064 & 0.34\\ 
  $\xi$ & 0.200 & 0.190 & -0.010 & 0.047 & 0.046 \\ 
  $\sigma$ & 1.000 & 1.098 & 0.098 &  0.127 & 0.081
\end{tabular}
        \caption{Asymptotic spatial dependence with $\lambda_2=0.5$}
        \label{tab:week2}
     \end{subtable}
     \end{table}

\clearpage

     \begin{table}[htbp]
     \ContinuedFloat
     \centering
    \begin{subtable}[h]{0.6\textwidth}
        \centering
        \begin{tabular}{rrrrrrr}
Parameter & True & Mean & Bias & RMSE & SE &  Coverage \\ 
  \hline
  $\lambda_1$ & 0.150 & 0.118  &  0.038  &  0.001 & 0.021 & 0.85\\ 
  $\lambda_2$ & 0.170 & 0.199  &  0.062  &  0.004 & 0.056 & 0.97\\ 
  $\lambda_3$ & 0.500 & 0.510  &  0.041  &  0.002 & 0.040 & 0.98\\ 
  $\lambda_4$ & 0.180 & 0.173  &  0.048  &  0.002 & 0.047 & 0.89\\ 
  $\rho_S$ & 0.400 & 0.406  &  0.149  &  0.022 & 0.150 & 0.97\\ 
  $\rho_T$ & 0.400 & 0.408  &  0.039  &  0.002 & 0.038 & 0.95\\ 
  $\xi$ & 0.200 & 0.191 & -0.009 & 0.032 & 0.031\\ 
  $\sigma$ & 1.000 & 1.063 & 0.063 & 0.088 &  0.061 
\end{tabular}
       \caption{Asymptotic temporal dependence with $\lambda_3=0.5$}
    \end{subtable}
    \vfill
    \begin{subtable}[h]{0.6\textwidth}
        \centering
       \begin{tabular}{rrrrrrr}
Parameter & True & Mean & Bias & RMSE & SE &  Coverage \\ 
  \hline
  $\lambda_1$ & 0.150 & 0.189  &  0.078  &  0.006 & 0.068 & 0.92\\ 
  $\lambda_2$ & 0.170 & 0.149  &  0.044  &  0.002 & 0.039 & 0.93\\ 
  $\lambda_3$ & 0.180 & 0.214  &  0.067  &  0.004 & 0.057 & 0.89 \\ 
  $\lambda_4$ & 0.500 & 0.448  &  0.131  &  0.017 & 0.120 & 0.90\\ 
  $\rho_S$ & 0.400 & 0.452  &  0.119  &  0.014 & 0.107 & 0.97\\ 
  $\rho_T$ & 0.400 & 0.414  &  0.068  &  0.005 & 0.066 & 0.94\\
  $\xi$ & 0.200 & 0.187 & -0.013 & 0.044 & 0.042 \\ 
  $\sigma$ & 1.000 & 1.092 & 0.092 & 0.133 & 0.097 \\
\end{tabular}
        \caption{Asymptotic independence with $\lambda_4=0.5$}
        \label{tab:week2}
     \end{subtable}
    \caption{Parameter estimates when asymptotic dependence is (a) spatio-temporal, (b) spatial, (c) temporal and (d) independent. The parameters $\lambda_j$ control the type of dependence; $\rho_S$ and $\rho_T$ are the spatial and temporal range parameters and $\xi$ and $\sigma$ are the marginal scale and shape parameters.  ``Coverage'' refers to the empirical coverage of 95\% bootstrap confidence intervals.}\label{t:sim}
\end{table}

\section{Analysis of extreme streamflow}\label{s:streamflow}

\begin{figure}
\caption{Scale parameter estimates across space}\label{f:streamflow:scale}\centering 
\includegraphics[width=0.8\textwidth,page=4, trim = {0 4cm 0 4cm}]{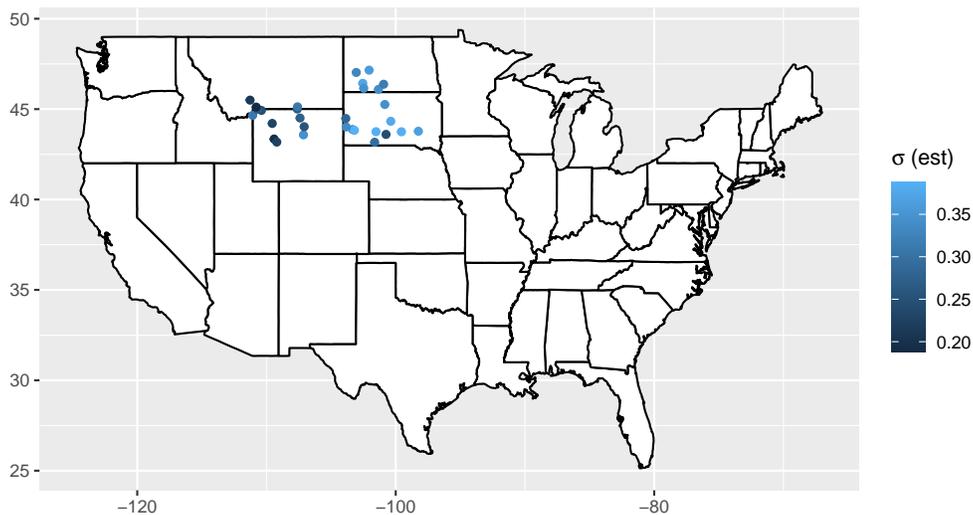}
\end{figure}

We illustrate our method using daily streamflow data downloaded from the USGS Hydro-Climatic Data Network \citep[HCDN;][]{lins2012usgs}.   We focus on the $m=30$ stations in the western/midwestern US;  these are mapped in Figure \ref{f:streamflow:scale} in USGS hydrologic unit codes region 10U and the $n=60$ years from 1965 to 2024. The spatiotemporal coordinates are projected onto the unit cube.  Because the vast majority of extreme values occur in the summer, we restrict our analysis to data from May to July. The data is downloaded using the \texttt{dataRetrieval} package in \texttt{R} \citep{dataRetrieval}. The objectives of the analysis are to estimate the marginal return levels at each site and test for asymptotic spatial and temporal dependence.

\begin{figure}\caption{QQ plot of marginal fits of streamflow data at thresholds $\tau = 0.5, 0.8, 0.9, 0.95$.}\label{f:streamflow:qq}\centering 
\includegraphics[page=2, width=0.4\textwidth]{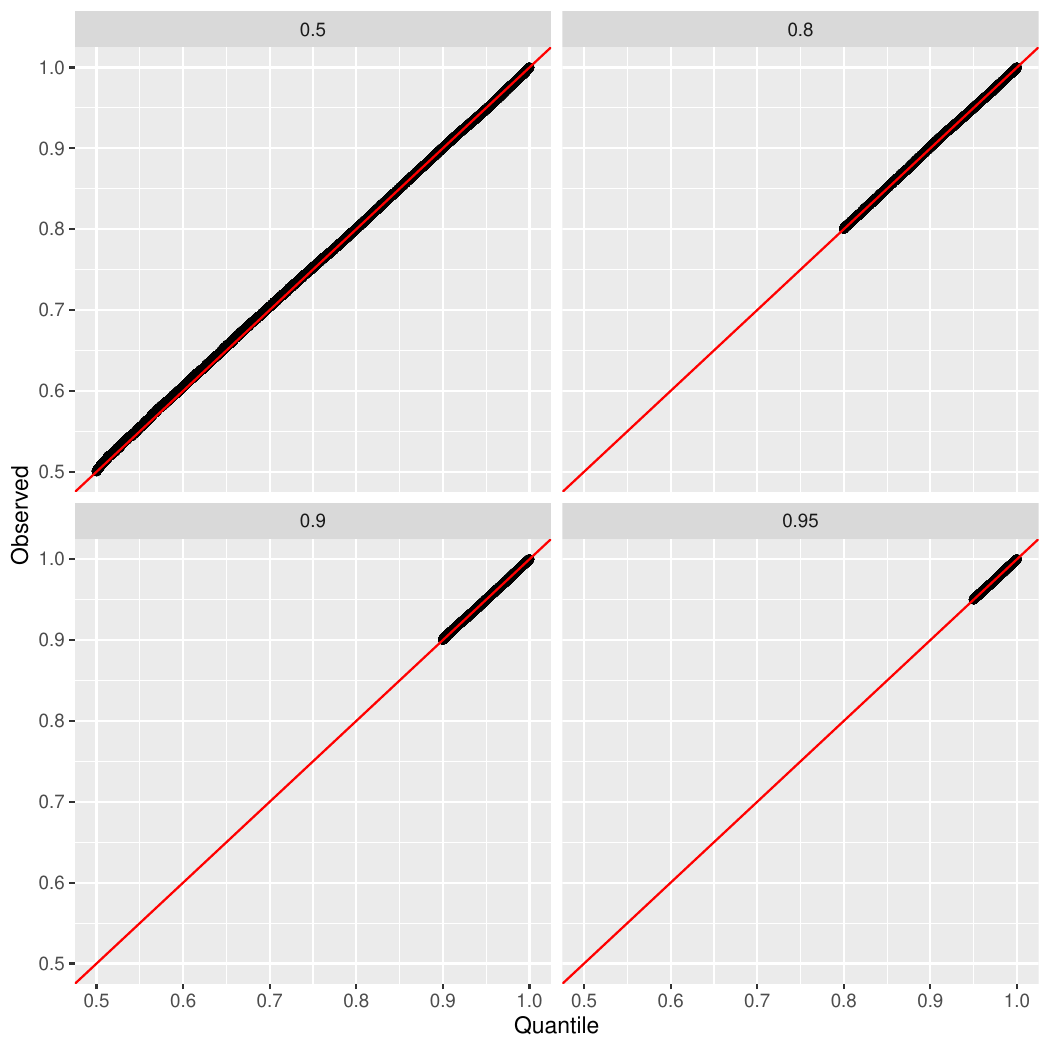}
\end{figure}

We first transform the raw data by taking its square root, which reduces the extreme heavy-tailedness for numerical stability. We then harmonize the data across stations by defining the response for year $i\in\{1,...,n\}$ for the station at spatial location $s$ and day of the year $t\in\{1,...,T\}$ as
\begin{equation}\label{e:Ytransformed}
Y_i(s,t) = \frac{Z_i(s,t)-q_{0.5}(s)}{q_{0.9}(s)-q_{0.1}(s)},
\end{equation}
where $Z_i(s,t)$ is the square root of the original measurement and $q_{\tau}(s)$ is the sample $\tau$ quantile of the observations at location $s$.  We then assume the marginal distribution is 
\begin{equation}\label{Y:gev}
Y_i(s,t)|Y_i(s,t) > Q_{\tau} \sim\mbox{GPD} \{ Q_{\tau},\sigma_i(s),\xi(s) \},
\end{equation}
where the threshold $Q_{\tau}$ is the empirical $\tau$ quantile of the data pooled across time at each location and $\sigma_i(s),\xi(s)$ are location-specific maximum likelihood estimates for the GPD parameter. Estimation is carried out using the \texttt{eva} package \citep{baderyan2020} based on the entire data, ignoring spatial and temporal dependence.  The QQ-plot in Figure \ref{f:streamflow:qq} shows the GPD model fits the data well for $\tau=0.8 $, which we use for the remainder of the analysis.

\begin{table}[]
    \centering
    \caption{Parameter Estimates and 95\% Bootstrap CIs}
    \label{tab:parest}
    \begin{tabular}{c|c|c}
        Parameter & Estimate & 95\% CI \\
        \hline
        $\lambda_1$ & 0.134 & (0.114, 0.174)\\
        $\lambda_2$ & 0.063 & (0.054, 0.139)\\
        $\lambda_3$ & 0.268 & (0.150, 0.302)\\
        $\lambda_4$ & 0.534 & (0.485, 0.625)\\
        $\rho_S$ & 0.261 & (0.214, 0.462)\\
        $\rho_T$ & 0.244 & (0.221, 0.341)\\
    \end{tabular}

\end{table}

The parameter estimates are reported in Table \ref{tab:parest}. 
Mixture parameter $\lambda_4$ was estimated as the largest, indicating that the data is asymptotically independent in both space and time. There is, however, significant weight in $\lambda_1$, which indicates a component with asymptotic dependence in space-time. Over 1000 non-parametric bootstrap samples, all $1000$ resulted in $\lambda_4$ being largest, indicating strong evidence that the data is asymptotically independent in space and time.

\begin{figure}\caption{Empirical $\chi$ statistics for pairs of stations plotted against distance for different temporal lags (right) for each station. The red line indicates $\chi$ statistics for fitted model.}\label{f:streamflow:chi}\centering 
\includegraphics[width=0.6\textwidth,page=3]{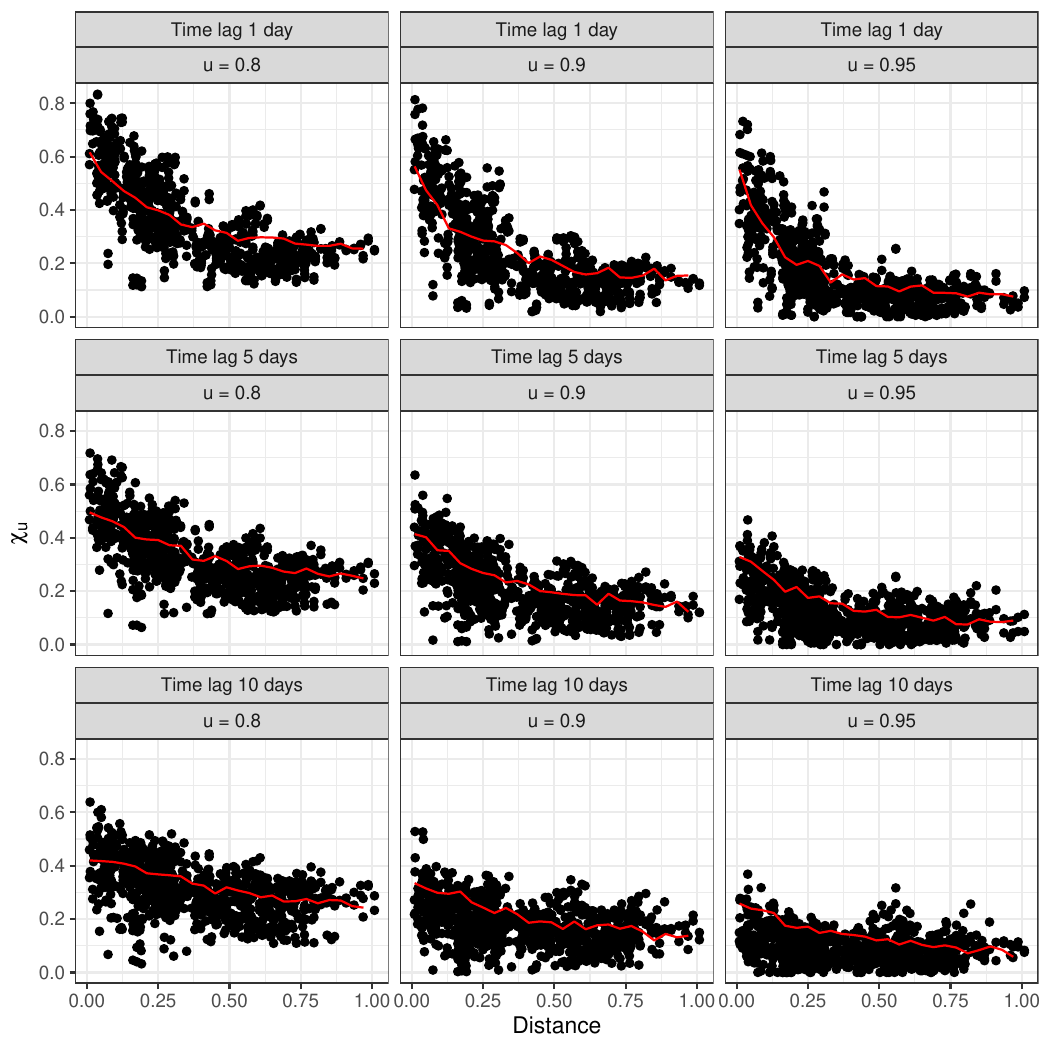}
\end{figure}

Figure $\ref{f:streamflow:chi}$ plots the empirical $\chi_u$ values against (normalized) distance in space, for $u = 0.8,0.9,0.95$ and time lags of 1, 5, and 10 days. The red line is a plot of approximate $\chi_u$ generated from our model at the fitted values. The modeled line matching the trend of the empirical $\chi_u$ from the data indicates that our model fit accurately reflects the pairwise asymptotic behavior of the real data across space-time. We can also see that $\chi_u \to 1-u$ as distance increases, indicating our model captures independence as spatial distance increases.

The estimated $\rho_S$ and $\rho_T$ parameters are $0.261$ and $0.244$ respectively, indicating that spatial dependence is limited to about 1/4 of our spatial domain and that temporal dependence is limited to about 15 days. The diagnostic plots in Figure \ref{f:streamflow:chi} show that the fitted spatiotemporal dependence functions matches the empirical $\chi$ coefficients and thus that the model fits the data well.  From the scale parameters plotted in Figure \ref{f:streamflow:scale}, we see that, across the spatial domain of interest, streamflow is higher for sites in the eastern portion of our spatial domain.

\section{Discussion}
In this paper, we proposed a process mixture model for spatial extremes whose spatio-temporal dependence is characterized by a convex combination of four random variables of differing asymptotic dependence. This model extends \citet{huser2019modeling} and \citet{majumder2024modeling} to include temporal dependence. Marginal parameters are estimated using standard maximum likelihood estimation, and dependence parameters are estimated using random forest regression.
We use this model to analyze daily streamflow data in the Western/Midwestern United States. We find that, over this domain, daily streamflow data is asymptotically independent over space and time. 

Future work will focus on improving estimation of model parameters. Although our results demonstrate that random forest regression is sufficient for estimation, estimating all parameters simultaneously may improve efficiency of uncertainty quantification, either with approximate likelihood approaches or a neural network approach. Additionally, estimating marginal parameters and dependence parameters simultaneously may improve efficiency for estimating marginal parameters as well. This model cannot make predictions at unobserved sites, both in space and in time; for applications like extreme streamflow, it may be useful to investigate whether it is possible to adapt this model to allow for prediction, so forecasts can be made to inform flood response policy. Finally, we hope to expand on the theoretical properties of the model. Although this paper derives $\chi$ for the model for different values of $\lambda$ in complete dependence and independence cases, it does not cover the general case where $R$ and $W$ are spatio-temporal processes.

\section*{Acknowledgements}

This work was supported by National Science Foundation grants DMS2152887 and CBET2151651.  The authors also thank Dr. Sankar Arumugam of North Carolina State University for providing and describing the streamflow data.

\bibliographystyle{apalike}
\bibliography{bib}
\clearpage
\section*{Appendix -- Proof of Theorem 1}

Let $X_1 = X(s_1,t_1), X_2 = X(s_2,t_2)$ be the value of the process at two spatio-temporal locations. 
Consider the joint survival function of $X_1, X_2$, $P(X_1 > y, X_2 > y)$, in 3 cases.

\subsection*{Case 1: $s_1 \ne s_2, t_1 = t_2$}

Let $t = t_1 = t_2$. We have $R_2(t_1) = R_2(t_2) = R_2$ so $X_1 = \lambda_1 R_1 + \lambda_2 R_2 + \lambda_3 R_3(s_1) + \lambda_4 W(s_1,t) $, $X_2 = \lambda_1 R_1 + \lambda_2 R_2 + \lambda_3 R_3(s_2) + \lambda_4 W(s_2,t) $.

Then, for $P(X_1 > y, X_2 > y)$, we have the following:

\begin{align*}
    P(X_1 > y, X_2 > y) & = P(\lambda_1 R_1 + \lambda_2 R_2 + \lambda_3 R_3(s_1) + \lambda_4 W(s_1,t) > y, \\ 
    & \qquad \lambda_1 R_1 + \lambda_2 R_2 + \lambda_3 R_3(s_2) + \lambda_4 W(s_2,t) > y)  \\
    & = E_{R_2}[ E_{R_1} \{P(\lambda_3 R_3(s_1) + \lambda_4 W(s_1,t) > y - \lambda_1 r_1 - \lambda_2 r_2, \\ 
    & \qquad \lambda_3 R_3(s_2) + \lambda_4 W(s_2,t) > y - \lambda_1 r_1 - \lambda_2 r_2| R_1 = r_1, R_2 = r_2)\} ] \\ 
   & = E_{R_2}[ E_{R_1} \{P(\lambda_3 R_3(s_1) + \lambda_4 W(s_1,t) > y - \lambda_1 r_1 - \lambda_2 r_2| R_1 = r_1, R_2 = r_2) \\ 
   & \qquad P(\lambda_3 R_3(s_2) + \lambda_4 W(s_2,t) > y - \lambda_1 r_1 - \lambda_2 r_2| R_1 = r_1, R_2 = r_2)\} ]
\end{align*}

\noindent where we used the independence of $R_3(s_1), R_3(s_2), W(s_1,t),$ and $W(s_2,t)$ to get the last line.

Since $\lambda_3 R_3(s_1) + \lambda_4 W(s_1,t)$ follows a hypoexponential distribution with parameters $\lambda_3, \lambda_4$, we know $P(\lambda_3 R_3(s_1) + \lambda_4 W(s_1,t) > x) = \frac{\lambda_3}{\lambda_3 - \lambda_4}\exp(-x/\lambda_3) - \frac{\lambda_4}{\lambda_3 - \lambda_4}\exp(-x/\lambda_4)$ if $x > 0$, and $1$ otherwise. Thus, equation (1) simplifies to:

\begin{align}\begin{split}
    & P(X_1 > y, X_2 > y) \\ 
    & = E_{R_1}[ E_{R_2} \{P(\lambda_3 R_3(s_1) + \lambda_4 W(s_1,t) > y - \lambda_1 r_1 - \lambda_2 r_2| R_1 = r_1, R_2 = r_2) \\ & \qquad \cdot P(\lambda_3 R_3(s_2) + \lambda_4 W(s_2,t) > y - \lambda_1 r_1 - \lambda_2 r_2| R_1 = r_1, R_2 = r_2) I(\lambda_1 r_1 + \lambda_2 r_2 > y)\} ] \\ & \qquad + E_{R_1}[ E_{R_2} \{P(\lambda_3 R_3(s_1) + \lambda_4 W(s_1,t) > y - \lambda_1 r_1 - \lambda_2 r_2| R_1 = r_1, R_2 = r_2) \\ & \qquad \cdot P(\lambda_3 R_3(s_2) + \lambda_4 W(s_2,t) > y - \lambda_1 r_1 - \lambda_2 r_2| R_1 = r_1, R_2 = r_2) I(\lambda_1 r_1 + \lambda_2 r_2 \le y)\} ] \\
    & = P(\lambda_1 R_1 + \lambda_2 R_2 > y) \\ & \qquad + E_{R_1}[ E_{R_2} \{P(\lambda_3 R_3(s_1) + \lambda_4 W(s_1,t) > y - \lambda_1 r_1 - \lambda_2 r_2| R_1 = r_1, R_2 = r_2) \\ & \qquad \cdot P(\lambda_3 R_3(s_2) + \lambda_4 W(s_2,t) > y - \lambda_1 r_1 - \lambda_2 r_2| R_1 = r_1, R_2 = r_2) I(\lambda_1 r_1 + \lambda_2 r_2 < y)\} ]
\end{split}\end{align}

Since $\lambda_1 R_1 + \lambda_2 R_2$ follows a hypoexponential distribution with parameters $\lambda_1, \lambda_2$, we know $P(\lambda_1 R_1 + \lambda_2 R_2 > y) = \frac{\lambda_1}{\lambda_1 - \lambda_2}\exp(-x/\lambda_1) - \frac{\lambda_2}{\lambda_1 - \lambda_2}\exp(-x/\lambda_2)$.

Next, consider the the second term in equation (2). 
\begin{align*}\begin{split}
    & E_{R_1} \big[E_{R_2} \{P(\lambda_3 R_3(s_1) + \lambda_4 W(s_1,t) > y - \lambda_1 r_1 - \lambda_2 r_2| R_1 = r_1, R_2 = r_2)^2 I(\lambda_1 r_1 + \lambda_2 r_2 < y)\}\big] \\
    & = E_{R_1} \left\{E_{R_2} \left(\left[\frac{\lambda_3}{\lambda_3-\lambda_4}\exp\left\{\frac{- (y - \lambda_1  r_1 - \lambda_2 r_2)} {\lambda_3} \right\} - \frac{\lambda_4}{\lambda_3-\lambda_4}\exp\left\{\frac{-(y - \lambda_1  r_1 - \lambda_2 r_2)}{\lambda_4} \right\}  \right]^2 I[\lambda_1 r_1 + \lambda_2 r_2 < y] \right) \right\} \\
    & = \int_0^{y/\lambda_1} \int_0^{(y-\lambda_1 r_1)/\lambda_2} \left[\left\{\frac{\lambda_3}{\lambda_3-\lambda_4}\right\}^2 \exp\left\{\frac{-2 (y - \lambda_1  r_1 - \lambda_2 r_2)}{\lambda_3} \right\} + \left\{\frac{\lambda_4}{\lambda_3-\lambda_4}\right\}^2\exp\left\{\frac{-2 (y - \lambda_1  r_1 - \lambda_2 r_2)}{\lambda_4} \right\} \right. \\
    & \left. \qquad - 2 \left\{\frac{\lambda_3 \lambda_4}{(\lambda_3-\lambda_4)^2}\right\} \exp\left\{-\left(\frac{1}{\lambda_3} + \frac{1}{\lambda_4}\right) (y - \lambda_1  r_1 - \lambda_2 r_2) \right\} \right]  e^{-r_2} dr_2 e^{-r_1} dr_1 \\
    & = I_1 + I_2 + I_3 
\end{split}\end{align*}

\noindent The first line is obtained by noting that $\lambda_3R_3(s_1) + \lambda_4W(s_1,t)$ and $\lambda_3R_3(s_2) + \lambda_4W(s_2,t)$ are identically distributed. The second line is obtained by substituting the hypoexponential survival function. 

We can separate this integral into three additive pieces, $I_1, I_2, I_3$, and handle them individually.

\begin{align*}
    I_1 & = \big\{\frac{\lambda_3}{\lambda_3-\lambda_4}\big\}^2 \int_0^{y/\lambda_1}  \exp\big\{\frac{-2 (y - \lambda_1  r_1)}{\lambda_3} \big\} \Big\{ \int_0^{(y-\lambda_1 r_1)/\lambda_2} \Big[\exp\big\{ \big(\frac{2\lambda_2}{\lambda_3} - 1\big) r_2 \big\}\Big] dr_2 \Big\} e^{-r_1} dr_1 \\
    & = \big\{\frac{\lambda_3}{\lambda_3-\lambda_4}\big\}^2 \frac{\lambda_3}{2\lambda_2 - \lambda_3} \int_0^{y/\lambda_1} \Big[ \exp\big\{\frac{-2 (y - \lambda_1  r_1)} {\lambda_3} \big\}  
    \exp\big\{ \big(\frac{2}{\lambda_3} - \frac{1}{\lambda_2}\big) (y - \lambda_1r_1) \big\} - 1\Big]   e^{-r_1} dr_1 \\
    & = \big\{\frac{\lambda_3}{\lambda_3-\lambda_4}\big\}^2 \frac{\lambda_3}{2\lambda_2 - \lambda_3}  \Big[\frac{\lambda_2}{\lambda_1 - \lambda_2}\big\{ \exp\big(-\frac{y}{\lambda_1}\big) - \exp\big(-\frac{y}{\lambda_2}\big)\big\} - \frac{\lambda_3}{2\lambda_1 - \lambda_3}\big\{ \exp\big(-\frac{y}{\lambda_1}\big) - \exp\big(-\frac{2y}{\lambda_3}\big) \big\} \Big]
\end{align*}

For brevity I skip the steps for integration for the second and third integrands in equation (3). The results are below:

\begin{align*}
    I_1 & = \int_0^{y/\lambda_1} \int_0^{(y-\lambda_1 r_1)/\lambda_2} [\{\frac{\lambda_3}{\lambda_3-\lambda_4}\}^2 \exp\{-2\lambda_3^{-1} (y - \lambda_1  r_1 - \lambda_2 r_2) \}] e^{-r_2} dr_2 e^{-r_1} dr_1 \\
    &  = (\frac{\lambda_3}{\lambda_3-\lambda_4})^2\frac{\lambda_3}{2\lambda_2 - \lambda_3}  [\frac{\lambda_2}{\lambda_1 - \lambda_2}\{ \exp(-\lambda_1^{-1}y) - \exp(-\lambda_2^{-1}y)\} - \frac{\lambda_3}{2\lambda_1 - \lambda_3}\{ \exp(-\lambda_1^{-1}y) - \exp(-2\lambda_3^{-1}y)\} ] \\
    I_2 & = \int_0^{y/\lambda_1} \int_0^{(y-\lambda_1 r_1)/\lambda_2} \{\frac{\lambda_4}{\lambda_3-\lambda_4}\}^2\exp\{-2\lambda_4^{-1} (y - \lambda_1  r_1 - \lambda_2 r_2) \}] e^{-r_2} dr_2 e^{-r_1} dr_1\\
    & = (\frac{\lambda_4}{\lambda_3-\lambda_4})^2\frac{\lambda_4}{2\lambda_2 - \lambda_4}  [\frac{\lambda_2}{\lambda_1 - \lambda_2}\{ \exp(-\lambda_1^{-1}y) - \exp(-\lambda_2^{-1}y)\} - \frac{\lambda_4}{2\lambda_1 - \lambda_4}\{ \exp(-\lambda_1^{-1}y) - \exp(-2\lambda_4^{-1}y)\} ]\\
    I_3 & = \int_0^{y/\lambda_1} \int_0^{(y-\lambda_1 r_1)/\lambda_2} -2 \frac{\lambda_3 \lambda_4}{(\lambda_3-\lambda_4)^2}\exp\{-(\lambda_3^{-1} + \lambda_4^{-1}) (y - \lambda_1  r_1 - \lambda_2 r_2) \}] e^{-r_2} dr_2 e^{-r_1} dr_1\\
    & = -2 \frac{\lambda_3 \lambda_4}{(\lambda_3-\lambda_4)^2} \frac{\lambda_3 \lambda_4}{\lambda_2(\lambda_3 + \lambda_4) - \lambda_3\lambda_4} [\frac{\lambda_2}{\lambda_1 - \lambda_2}\{ \exp(-\lambda_1^{-1}y) - \exp(-\lambda_2^{-1}y)\} \\
    & \qquad- \frac{\lambda_3\lambda_4}{\lambda_1(\lambda_3 + \lambda_4) - \lambda_3\lambda_4} \{ \exp(-\lambda_1^{-1}y) - \exp(-(\lambda_3^{-1} + \lambda_4^{-1}) y)\}] 
\end{align*}

Gathering terms and simplifying gives $P(X_1 > y, X_2 > y) = c_1e^{-y/\lambda_1} + c_2e^{-y/\lambda_2} + c_3 e^{-2y/\lambda_3} + c_4 e^{-2y/\lambda_4} + c_5e^{-(y/\lambda_3 + y/\lambda_4)} $ where

\begin{align*}
    c_1 & =  \frac{\lambda_1}{(\lambda_1 - \lambda_2)} + \frac{\lambda_1\lambda_3^3}{(\lambda_1 - \lambda_2)(\lambda_3 - \lambda_4)^2(2\lambda_1 - \lambda_3)}  
    + \frac{\lambda_1\lambda_4^3}{(\lambda_1 - \lambda_2)(\lambda_3 - \lambda_4)^2(2\lambda_1 - \lambda_4)} 
    \\ & \qquad - \frac{2\lambda_1\lambda_3^2\lambda_4^2}{(\lambda_1 - \lambda_2)(\lambda_3 - \lambda_4)^2(\lambda_1\lambda_3 + \lambda_1\lambda_4 - \lambda_3\lambda_4)}\\
    c_2 & =   -\frac{\lambda_2}{(\lambda_1 - \lambda_2)} - \frac{\lambda_2\lambda_3^3}{(\lambda_1 - \lambda_2)(\lambda_3 - \lambda_4)^2(2\lambda_2 - \lambda_3)} - \frac{\lambda_2\lambda_4^3}{(\lambda_1 - \lambda_2)(\lambda_3 - \lambda_4)^2(2\lambda_2 - \lambda_4)} \\ & \qquad + \frac{2\lambda_2\lambda_3^2\lambda_4^2}{(\lambda_1 - \lambda_2)(\lambda_3 - \lambda_4)^2(\lambda_2\lambda_3 + \lambda_2\lambda_4 - \lambda_3\lambda_4)}\\
    c_3 & =  \frac{\lambda_3^4}{(\lambda_3 - \lambda_4)^2 (2\lambda_1 - \lambda_3) (2\lambda_2 - \lambda_3)}\\
    c_4 & =  \frac{\lambda_4^4}{(\lambda_3 - \lambda_4)^2 (2\lambda_1 - \lambda_4) (2\lambda_2 - \lambda_4)}\\
    c_5 & = \frac{-2\lambda_3^3\lambda_4^3}{(\lambda_3 - \lambda_4)^2 (\lambda_2\lambda_3 + \lambda_2\lambda_4 - \lambda_3\lambda_4) (\lambda_1\lambda_3 + \lambda_1\lambda_4 - \lambda_3\lambda_4)}\\    
\end{align*}

Now, we have an expression for $P(X_1 > y, X_2 > y)$. Additionally, since $X_1$ follows a 4-parameter hypoexponential distribution, we have:

\[P(X_1 > y) = \sum_{i=1}^4 \exp(-\lambda_i^{-1} y)\prod_{1 \le j \le 4, j \ne i}\Big(\frac{\lambda_i}{ \lambda_i - \lambda_j}\Big) \]

Then, if the index of the largest $\lambda_i$, $\underset{i}{\operatorname{argmax}}\lambda_i = 3$ or $4$, $\lim_{y \to \infty} P(X_1 > y, X_2 > y) / P(X_1 > y) = 0$ so $X_1$ and $X_2$ will be asymptotically independent. Otherwise, if $\underset{i}{\operatorname{argmax}}\lambda_i = 1$ or $2$, then $X_1$ and $X_2$ will be asymptotically dependent and the limit $\chi(X_1, X_2)$ will be $c_j \prod_{1 \le l \le 4, l \ne l}(\frac{\lambda_j}{ \lambda_j - \lambda_l})^{-1}$ where $j = \underset{i}{\operatorname{argmax}}\lambda_i$ and $c_i$ is as defined above. 

This result holds if $c_1, c_2$ are non-zero. To verify this, I expand the numerator terms of $c_1, c_2$ and find the roots of the resulting polynomial in Wolfram Alpha. None of the roots occurred in the restricted domain we proposed for $\lambda_i$, so we conclude this is not a concern.


\subsection*{Case 2: $s_1 = s_2, t_1 \ne t_2$}

This case mirrors the proof above, so I will not include it for brevity. 

We find that if the index of the largest $\lambda_i$, $\underset{i}{\operatorname{argmax}}\lambda_i = 2$ or $4$, $\lim_{y \to \infty} P(X_1 > y, X_2 > y) / P(X_1 > y) = 0$ so $X_1$ and $X_2$ will be asymptotically independent. Otherwise, if $\underset{i}{\operatorname{argmax}}\lambda_i = 1$ or $3$, then $X_1$ and $X_2$ will be asymptotically dependent and the limit $\chi(X_1, X_2)$ will be $c_j \prod_{1 \le l \le 4, l \ne l}(\frac{\lambda_j}{ \lambda_j - \lambda_l})^{-1}$ where $j = \underset{i}{\operatorname{argmax}}\lambda_i$ and $c_i$ is as defined above. 

\subsection*{Case 3: $s_1 \ne s_2, t_1 \ne t_2$}

We have $X_1 = \lambda_1 R_1 + \lambda_2 R_2(t_1) + \lambda_3 R_3(s_1) + \lambda_4 W(s_1,t_1) $, $X_2 = \lambda_1 R_1 + \lambda_2 R_2(t_2) + \lambda_3 R_3(s_2) + \lambda_4 W(s_2,t_2) $.

Then, we have the following:

\begin{align*}
    P(X_1 > y, X_2 > y) & = P(\lambda_1 R_1 + \lambda_2 R_2 + \lambda_3 R_3(s_1) + \lambda_4 W(s_1,t) > y, \lambda_1 R_1 + \lambda_2 R_2 + \lambda_3 R_3(s_2) \\ 
    & \qquad  + \lambda_4 W(s_2,t) > y) \\
    & = E_{R_1} \{P(\lambda_2 R_2(t_1) + \lambda_3 R_3(s_1) + \lambda_4 W(s_1,t_1) > y - \lambda_1 r_1, \lambda_2 R_2(t_2) + \lambda_3 R_3(s_2) \\ & \qquad + \lambda_4 W(s_2,t_2) > y - \lambda_1 r_1| R_1 = r_1)\} \\
    & = P(R_1 > y/\lambda_1) + \int_{0}^{y/\lambda_1} P(\lambda_2 R_2 + \lambda_3 R_3 + \lambda_4 W > y - \lambda_1 r)^2 e^{-r} dr
\end{align*}

Since $R_1$ follows a standard exponential distribution, $P(R_1 > y/\lambda_1) = \exp(-y/\lambda_1)$ Since $\lambda_2 R_2 + \lambda_3 R_3 + \lambda_4 W$ follows a hypoexponential distribution with parameters $\lambda_2, \lambda_3, \lambda_4$, we know:

\[P(\lambda_2 R_2 + \lambda_3 R_3 + \lambda_4 W > x) = \Bigg\{ \frac{\lambda_2^2 \exp(-y/\lambda_2)}{(\lambda_2 - \lambda_3)(\lambda_2 - \lambda_4)} - \frac{\lambda_3^2 \exp(-y/\lambda_3)}{(\lambda_2^ - \lambda_3)(\lambda_3 - \lambda_4)} 
+ \frac{\lambda_4^2 \exp(-y/\lambda_4)}{(\lambda_2 - \lambda_4)(\lambda_3 - \lambda_4)}\Bigg\}\]
 $\qquad$ if $x > 0$, and $1$ otherwise.

The expression in the integral can be expanded as follows:

\begin{align*}
    & \int_{0}^{y/\lambda_1} P(\lambda_2 R_2 + \lambda_3 R_3 + \lambda_4 W > y - \lambda_1 r)^2 e^{-r} dr \\ 
    & = \int_{0}^{y/\lambda_1}  \Bigg[ \frac{\lambda_2^2 \exp\{-\lambda_2^{-1}(y-\lambda_1 r)\}}{(\lambda_2 - \lambda_3)(\lambda_2 - \lambda_4)} - \frac{\lambda_3^2 \exp\{-\lambda_3^{-1}(y-\lambda_1 r)\}}{(\lambda_2 - \lambda_3)(\lambda_3 - \lambda_4)} + \frac{\lambda_4^2 \exp\{-\lambda_4^{-1}(y-\lambda_1 r)\}}{(\lambda_2 - \lambda_4)(\lambda_3 - \lambda_4)}\Bigg]^2 e^{-r} dr \\
        & = \int_{0}^{y/\lambda_1} \Bigg[ \frac{\lambda_2^4 \exp\{-2\lambda_2^{-1}(y-\lambda_1 r)\}}{(\lambda_2 - \lambda_3)^2(\lambda_2 - \lambda_4)^2} + \frac{-\lambda_3^4 \exp\{-2\lambda_3^{-1}(y-\lambda_1 r)\}}{(\lambda_2 - \lambda_3)^2(\lambda_3 - \lambda_4)^2} + \frac{-\lambda_4^4 \exp\{-2\lambda_4^{-1}(y-\lambda_1 r)\}}{(\lambda_2 - \lambda_4)^2(\lambda_3 - \lambda_4)^2} \\ 
    & \qquad - \frac{2\lambda_2^2\lambda_3^2 \exp\{-(y-\lambda_1 r)(\lambda_2^{-1} + \lambda_3^{-1})\}}{(\lambda_2 - \lambda_3)^2(\lambda_2 - \lambda_4)(\lambda_3 - \lambda_4)} - \frac{2\lambda_3^2\lambda_4^2 \exp\{-(y-\lambda_1 r)(\lambda_3^{-1} + \lambda_4^{-1})\}}{(\lambda_3 - \lambda_4)^2(\lambda_2 - \lambda_3)(\lambda_2 - \lambda_4)} \\
    & \qquad + \frac{2\lambda_2^2\lambda_4^2 \exp\{-(y-\lambda_1 r)(\lambda_2^{-1} + \lambda_4^{-1})\}}{(\lambda_2 - \lambda_4)^2(\lambda_2 - \lambda_3)(\lambda_3 - \lambda_4)} \Bigg] e^{-r} dr \\
    & = J_1 + J_2 + J_3 + J_4 + J_5 + J_6
\end{align*} 

We can split this integral additively and evaluate each piece separately.

\begin{align*}
    J_1 & = \int_{0}^{y/\lambda_1} \frac{\lambda_2^4 \exp\{-2\lambda_2^{-1}(y-\lambda_1 r)\}}{(\lambda_2 - \lambda_3)^2(\lambda_2 - \lambda_4)^2}  e^{-r} dr \\
    &  = \frac{\lambda_2^4 \exp(-2\lambda_2^{-1}y) }{(\lambda_2 - \lambda_3)^2(\lambda_2 - \lambda_4)^2} \int_{0}^{y/\lambda_1} \exp\{(\frac{2\lambda_1}{\lambda_2} - 1) r\}  dr \\
    & = \frac{\lambda_2^4 \exp(-2\lambda_2^{-1}y) }{(\lambda_2 - \lambda_3)^2(\lambda_2 - \lambda_4)^2} \Big[\frac{\exp\{(2\lambda_2^{-1} - \lambda_1^{-1}) y\} - 1}{\frac{2\lambda_1}{\lambda_2} - 1}\Big] \\
    & = \frac{\lambda_2^5 \{\exp(-\lambda_1^{-1}y) - \exp(-2\lambda_2^{-1}y)\} }{(\lambda_2 - \lambda_3)^2(\lambda_2 - \lambda_4)^2(2\lambda_1 - \lambda_2)} 
\end{align*}

For brevity I skip the steps for integration for the other five terms. 
\begin{align*}
    J_2 & = \int_{0}^{y/\lambda_1} \frac{\lambda_3^4 \exp\{-2(y-\lambda_1 r)/\lambda_3\}}{(\lambda_2 - \lambda_3)^2(\lambda_3 - \lambda_4)^2}  e^{-r} dr = \frac{\lambda_3^5 \{\exp(-y/\lambda_1) - \exp(-2y/\lambda_3)\} }{(\lambda_2 - \lambda_3)^2(\lambda_3 - \lambda_4)^2(2\lambda_1 - \lambda_2)}\\
    J_3 & = \int_{0}^{y/\lambda_1} \frac{\lambda_4^4 \exp\{-2(y-\lambda_1 r)/\lambda_4\}}{(\lambda_2 - \lambda_4)^2(\lambda_3 - \lambda_4)^2}  e^{-r} dr = \frac{\lambda_4^5 \{\exp(-y/\lambda_1) - \exp(-2y/\lambda_4)\} }{(\lambda_2 - \lambda_4)^2(\lambda_3 - \lambda_4)^2(2\lambda_1 - \lambda_4)}\\
    J_4 & =\int_{0}^{y/\lambda_1} \frac{2\lambda_2^2\lambda_3^2 \exp\{-(y-\lambda_1 r)(\lambda_2^{-1} + \lambda_3^{-1})\}}{(\lambda_2 - \lambda_3)^2(\lambda_2 - \lambda_4)(\lambda_3 - \lambda_4)} e^{-r}dr \\
    & = \frac{2\lambda_2^3\lambda_3^3 \big[\exp\{-y/\lambda_1\} - \exp\{-y(\lambda_2^{-1} + \lambda_3^{-1}) \}  \big]}{(\lambda_2 - \lambda_3)^2(\lambda_2 - \lambda_4)(\lambda_3 - \lambda_4) (\lambda_1\lambda_2 + \lambda_1\lambda_3 - \lambda_2\lambda_3)}\\
    J_5 & = \int_{0}^{y/\lambda_1} \frac{2\lambda_2^2\lambda_4^2 \exp\{-(y-\lambda_1 r)(\lambda_2^{-1} + \lambda_4^{-1})\}}{(\lambda_2 - \lambda_4)^2(\lambda_2 - \lambda_2)(\lambda_3 - \lambda_4)} e^{-r}dr \\
    & = \frac{2\lambda_2^3\lambda_4^3 \big[\exp\{-y/\lambda_1\} - \exp\{-y(\lambda_2^{-1} + \lambda_4^{-1}) \}  \big]}{(\lambda_2 - \lambda_4)^2(\lambda_2 - \lambda_3)(\lambda_3 - \lambda_4) (\lambda_1\lambda_2 + \lambda_1\lambda_4 - \lambda_2\lambda_4)} \\
    J_6 & =  \int_{0}^{y/\lambda_1} \frac{2\lambda_3^2\lambda_4^2 \exp\{-(y-\lambda_1 r)(\lambda_3^{-1} + \lambda_4^{-1})\}}{(\lambda_3 - \lambda_4)^2(\lambda_2 - \lambda_3)(\lambda_2 - \lambda_4)} e^{-r}dr \\
    & = \frac{2\lambda_3^3\lambda_4^3 \big[\exp\{-y/\lambda_1\} - \exp\{-y(\lambda_3^{-1} + \lambda_4^{-1}) \}  \big]}{(\lambda_3 - \lambda_4)^2(\lambda_2 - \lambda_3)(\lambda_2 - \lambda_4) (\lambda_1\lambda_3 + \lambda_1\lambda_4 - \lambda_3\lambda_4)}
\end{align*}

We can see the joint survival probability is $c_1 \exp(-y/\lambda_1) + c_2 \exp(-2y/\lambda_2) + c_3 \exp(-2y/\lambda_3) + c_4 \exp(-2y/\lambda_4) + c_5 \exp\{-y(\lambda_2^{-1} + \lambda_3^{-1})\} + c_6 \exp\{-y(\lambda_2^{-1} + \lambda_4^{-1})\} + c_7 \exp\{-y(\lambda_3^{-1} + \lambda_4^{-1})\}$ where 
\begin{align*}
    c_1 & = -\sum_{i=2}^7 c_i\\
    c_2 & = \frac{-\lambda_2^5}{(\lambda_2 - \lambda_3)^2(\lambda_2 - \lambda_4)^2(2\lambda_1 - \lambda_2)}\\
    c_3 & = \frac{-\lambda_3^5}{(\lambda_2 - \lambda_3)^2(\lambda_3 - \lambda_4)^2(2\lambda_1 - \lambda_3)}\\
    c_4 & = \frac{-\lambda_4^5}{(\lambda_2 - \lambda_4)^2(\lambda_3 - \lambda_4)^2(2\lambda_1 - \lambda_4)}\\
    c_5 & = \frac{2\lambda_2^3\lambda_3^3 }{(\lambda_2 - \lambda_3)^2(\lambda_2 - \lambda_4)(\lambda_3 - \lambda_4) (\lambda_1\lambda_2 + \lambda_1\lambda_3 - \lambda_2\lambda_3)}\\
    c_6 & = \frac{-2\lambda_2^3\lambda_4^3 }{(\lambda_2 - \lambda_4)^2(\lambda_2 - \lambda_3)(\lambda_3 - \lambda_4) (\lambda_1\lambda_2 + \lambda_1\lambda_4 - \lambda_2\lambda_4)}\\
    c_7 & = \frac{2\lambda_3^3\lambda_4^3 }{(\lambda_3 - \lambda_4)^2(\lambda_2 - \lambda_3)(\lambda_2 - \lambda_4) (\lambda_1\lambda_3 + \lambda_1\lambda_4 - \lambda_3\lambda_4)}
\end{align*}


Additionally, since $X_1$ follows a four-parameter hypoexponential distribution, we have:

\[P(X_1 > y) = \sum_{i=1}^4 \exp(-\lambda_i^{-1} y)\prod_{1 \le j \le 4, j \ne i}(\frac{\lambda_i}{ \lambda_i - \lambda_j}) \]

Then, it is clear that $\lim_{y \to \infty} P(X_1 > y, X_2 > y) / P(X_1 > y) = 0$ only when ${\operatorname{argmax}}_{i} \lambda_i \ne 1$, so it will be asymptotically independent. Otherwise, $X_1$ and $X_2$ will be asymptotically dependent, and the limit $\chi$ will be:
\[\frac{c_1}{\prod_{2 \le j \le 4}(\frac{\lambda_i}{ \lambda_1 - \lambda_j})}\]
the ratio of coefficients of the $\exp(-y/\lambda_1)$ terms derived in the previous steps.


\end{document}